# All-atom simulations reveal how single point mutations promote serpin misfolding

**Authors:** F. Wang[†], S. Orioli[†], A. Ianeselli, G. Spagnolli,  S, a Beccara, A. Gershenson[*], P. Faccioli[*] & P.L. Wintrode[*]

*To whom correspondence should be addressed.

[†]These authors contributed equally to this work.

**Abstract**:

Protein misfolding is implicated in many diseases, including the serpinopathies. For the canonical inhibitory serpin α₁-antitrypsin (A1AT), mutations can result in protein deficiencies leading to lung disease, and misfolded mutants can accumulate in hepatocytes leading to liver disease. Using all-atom simulations based on the recently developed Bias Functional algorithm we elucidate how wild-type A1AT folds and how the disease-associated S (Glu264Val) and Z (Glu342Lys) mutations lead to misfolding. The deleterious Z mutation disrupts folding at an early stage, while the relatively benign S mutant shows late stage minor misfolding. A number of suppressor mutations ameliorate the effects of the Z mutation and simulations on these mutants help to elucidate the relative roles of steric clashes and electrostatic interactions in Z misfolding. These results demonstrate a striking correlation between atomistic events and disease severity and shine light on the mechanisms driving chains away from their correct folding routes.

Understanding how mutations alter protein misfolding propensities and the physico-chemical mechanisms underlying this shift is key to clarifying the molecular basis of many diseases. One set of relatively common protein misfolding diseases, the serpinopathies, arise when mutations in inhibitory serpins lead to misfolding, thus reducing the secreted levels of these important protease inhibitors (1). Mutations in the canonical secretory serpin α₁-antitrypsin (A1AT) result in the most common serpinopathies, the A1AT deficiencies. In these deficiencies, low circulating A1AT levels dysregulate leukocyte serine proteases resulting in lung disease which can be slowed but not halted by A1AT augmentation therapy (2). Extremely pathogenic A1AT mutations, such as Z (Glu342Lys), can lead to both lung disease, due to loss of function, and liver disease, due to A1AT accumulation in the endoplasmic reticulum (ER) of hepatocytes, which generate most of the circulating A1AT. With the exception of liver transplants, there are no effective treatments for A1AT associated liver disease (3).

*In vitro,* the pathogenic A1AT Z mutant folds very slowly, spending hours in at least one partially folded intermediate state (4). Similarly, Z secretion from cells is slow and, while some Z species are targeted for degradation (5–7), misfolded Z accumulates in the ER where it can polymerize (8). Despite numerous experimental studies (9–13), little is known about the structure of misfolded species for any A1AT disease-associated mutant, hindering efforts to either rescue the folding of these species or to target them for degradation.

Molecular dynamics (MD) simulations offer an attractive approach to study protein folding and misfolding, as they can in principle reveal folding pathways and intermediates in atomistic detail. To date the application of all-atom MD simulations to investigate protein folding and misfolding has been limited to small single-domain proteins, with relatively short folding times. In particular, recent developments such as the Anton special-purpose supercomputer (14) and the massively distributed folding@home project (15) have made it possible to generate *in silico* several reversible folding/unfolding events for a number of small globular proteins (< 100 amino acids) with folding times up to the ms range. These studies have demonstrated that current all-atom force fields in explicit solvent can lead to the correct native states of proteins and predict with good accuracy their folding kinetics. Unfortunately, most biologically relevant proteins are much larger than 100 amino acids and have folding times as long as seconds and beyond. In particular, A1AT and other serpins contain approximately 400 amino acids and fold over tens of minutes (9, 12, 13). Due to their large size and slow folding kinetics, simulating serpin folding with conventional MD simulations is not feasible, even using the most powerful available supercomputers.

While the folding of a single large protein is generally much too slow to simulate using standard MD, this is chiefly due to the exponentially long time spent in local free energy minima. However, single molecule studies have shown that, even for slow folding proteins, actual transitions between unfolded and folded conformations are rapid (typically on the microsecond time scale) and similar to those of small fast folding proteins (16). Therefore, if we can preferentially sample only the *reactive* parts of folding trajectories, it becomes feasible to accurately compute the folding pathways of even large proteins (17–19). In this work, we rely on a recently developed variational method called the Bias Functional (BF) approach (see Methods) to achieve this goal and sample transition paths. We use this scheme to characterize the folding and misfolding of wild-type (WT) A1AT, the pathological Z mutant and the relatively benign S (Glu264Val) mutant (1), starting from several fully denatured conformations. The results of these all-atom simulations provide testable, atomistic models for how WT A1AT folds, how disease associated mutants misfold and how suppressor mutations can rescue misfolding. These results also shine light on the connections between single-point mutations and the pathogenicity of misfolding prone proteins, and provide physical mechanisms responsible for misfolding phenotypes.

**Methods:**

The computational limitations of conventional MD have motivated the development of more sophisticated algorithms and approximations to study rare biomolecular transitions. Among these, well-tempered meta-dynamics has proved to be a powerful tool to profile the potential of mean-force of the slowest collective variables in a system (20). Conversely, transition paths sampling (17) offers a framework to directly sample reaction pathways, while transition interface sampling (21) and milestoning (22) enable the investigation of reaction kinetics. Unfortunately, to our knowledge, none of these powerful equilibrium and non-equilibrium methods have been successfully applied to reactions as complex and slow as serpin folding and misfolding.

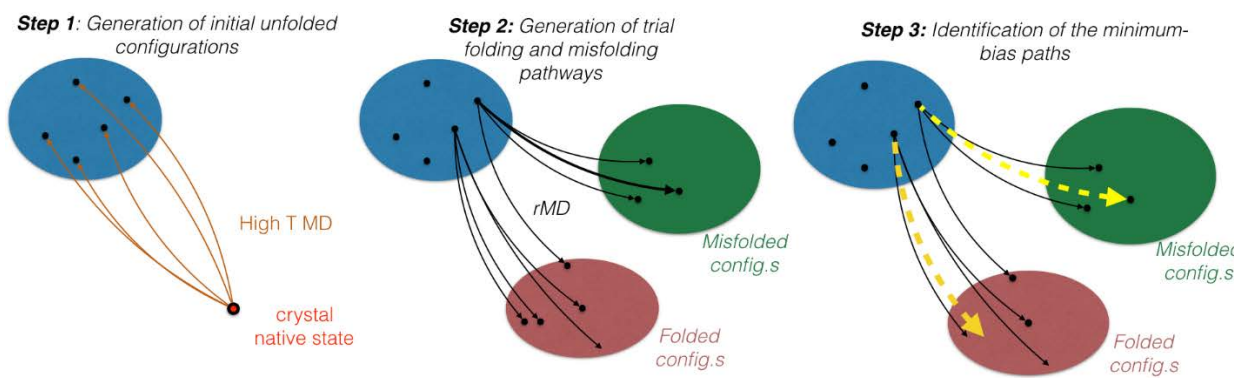

**Figure 1.** Schematic representation of the BF algorithm. Black solid lines are trial ratche-and-pawl (rMD) trajectories obtained starting from conformations in the unfolded set U (blue) generated in Step 1. The dashed yellow line is the minimum bias path, for which the value of the bias functional T given in Equation (2) is least.

**The Bias Functional (BF) Approach.** In the BF approach (23) a set of variationally optimized trajectories connecting an ensemble of fully denatured serpin conformations to folded and misfolded conformations is obtained through the following three-step procedure, schematically represented in Figure 1.

*Step 1: Generation of initial denatured conformations.* We sampled an ensemble of fully unfolded conformations (with a fraction of native contacts, $Q < 0.1$) by performing thermal unfolding MD simulations at 1600 K for 10 ns with an integration time of 1 fs starting from the WT native structure. Thermal unfolding was followed by a 10 ns relaxation at room temperature, T=300 K.

*Step 2: Generation of trial folding and misfolding pathways*: From each unfolded conformation $X_i$ obtained at step 1, we initiated 12 ratchet-and-pawl MD (rMD) simulations. In rMD, the standard MD at room temperature $T$ is modified by adding an unphysical history-dependent biasing force $F_{rMD}^i$ defined as follows (24, 25).

$$\mathbf{F}_{rMD}^i(X,t) = \begin{cases} -k_R \ \nabla^i z(X)(z[X(t)] - z_m(t)), & \text{for} \quad z[X(t)] > z_m(t) \\ 0, & \text{for} \quad z[X(t)] \leq z_m(t). \end{cases} \quad (1)$$

where, for a given BF folding trajectory, $X(t)$ is the A1AT conformation at time $t$. In this equation, $z(X)$ is a collective variable that measures the distance between the protein's *instantaneous* contact map $C_{ij}(X(t))$ and a reference contact map, calculated from the native state crystal structure $X^{native}$, i.e. ., $z(X) = \sum_{i<j} [ \ C_{ij}(X) - C_{ij}(X^{native}) ]^2$. $C_{ij}$ is a function of the distance between atom i and j, which interpolates between 0 and 1 and provides a continuum representation of the contact map $C_{ij}(X) = [1-(r_{ij}/r_0)^6]/[1-(r_{ij}/r_0)^{10}]$, where $r_0=0.75$ nm. The variable $z_m(t)$ in Equation 1 denotes the minimum value assumed by the collective coordinate $z$ up to time t along the given rMD trajectory. Thus, the biasing force (Eqn. 1) remains latent (equals 0) any time an rMD step leads to no change or to an increase of the overlap between the instantaneous and the native contact maps, i.e. if $z(X(t+\Delta t)) \leq z_m(t)$. On the other hand, the bias switches on whenever an elementary integration step leads to a decrease of the overlap with the native state, i.e. for $z(X(t+\Delta t)) > z_m(t)$.

*Step 3: Identification of the least-biased transition pathways.* The rMD paths generated in Step 2 are all affected by the presence of the unphysical biasing force $F_{rMD}^i$. However, it can be rigorously shown (23) that the trajectories in this trial path ensemble with the largest probability to occur in an *unbiased* Langevin simulation are those with the least value of the so-called bias functional:

$$T[X] = \int_0^t d\tau \sum_{i=1}^N \frac{1}{\gamma_i m_i} \ |\mathbf{F}_{rMD}^i(X;\tau)|^2. \quad (2)$$

where $\gamma_i$ and $m_i$ denote the viscosity and mass of the *i*-th atom, respectively.

Thus, for each denatured conformation $X_i$ we retain only one minimum-bias trajectory and discard all other trial rMD paths. This procedure keeps to a minimum the systematic errors introduced by the biasing force. It is important to recall that in rMD any time the chain progresses towards conformations with increased numbers of native contacts, the biasing force is not active. Furthermore, the strength of the biasing force is chosen in order to ensure that the modulus of the total biasing force is always at least two orders of magnitude smaller than that of the physical force. Consequently, the partially folded and misfolded structures obtained in our simulations are sensitive to the details of the underlying force field.

On the other hand, the presence of a non-physical biasing force could in principle promote the early formation of some local native contacts. While the BF variational principle is supposed to minimize these artifacts, some residual systematic errors may persist.

**Force Fields.** MD and rMD simulations were performed using the AMBER ff99SB force field (26), within the Generalized Born scheme implemented in GROMACS 4.5.2 (27). In general, the use of an implicit solvent model in protein folding MD simulations is considered a rather strong approximation. However, we note that previous applications of the BF approach and closely related variants used the same implicit solvent force field and produced results which compared well with those from MD simulations in explicit solvent (28, 29) and with experiments (30). Trial trajectories were generated by running 30 ns of rMD simulations at room temperature, 300 K, with an integration time of 1 fs. The parameter $k_R$ in Equation 1, setting the strength of the biasing force, was 0.02 kJ/mol.

**Simulations Details.** The X-ray crystal structure of active, native wildtype (WT) $\alpha_1$-antitrypsin (A1AT) from Elliott and co-workers (31) (PDB: 1QLP) was used as the target structure for WT BF folding simulations. To our knowledge, the first 22 amino acids are disordered in all published A1AT structures. Thus, the sequences for the folding simulations start at Phe23. We note that, from a kinetic point of view, this active native state is only metastable. Indeed, A1AT spontaneously performs an irreversible conformational reaction into a fully stable, yet biologically latent structure with a mean-first-passage time on the order of days to weeks. The latency transition was previously characterized in atomistic detail using a variant of the BF algorithm (32).

For all A1AT mutants, single point mutations were introduced using the mutator function in VMD (33). After we began our simulations, an X-ray crystal structure of the Z mutant was published, PDB: 5IO1 (34). There are no significant differences between the WT and Z active, native structures. A structural alignment of our target Z structure with the published structure shows a root mean square deviation (RMSD) of 0.6 Å and the native contact maps are identical.

For simulations of WT folding we generated 12 independent fully denatured conformations by thermal unfolding at 1600 K. From each such condition we produced 12 trial rMD trajectories and extracted the Least Biased Trajectory (LBT) using the minimum bias condition, i.e. selecting the path with the least BF functional, defined in Eqn. 2. Typically, for such a number of trial trajectories the requirements for convergence described in Ref. (23) are satisfied. For Z, S, and other A1AT variants, we also began each BF simulation with 12 independent unfolded conformations. Twelve LBTs of A1AT folding from 12 independent unfolded states can be generated in roughly 1 week using 32 cores on a standard computing cluster. This procedure was repeated multiple times yielding the total number of LBTs for each A1AT variant given in Figure 4A.

Even under the effect of the rMD biasing force, a residual degree of frustration persists. As a result, only 2% of our short WT A1AT LBTs reach the fully folded structure within the simulated time window. The majority of LBTs reach a fraction native contacts of at least 0.4, corresponding to the formation of the local foldons (Fig. 2, stage 2A). 45% of all WT LBTs reach stage 4 or 5 shown in Figure 2A, where the central β-sheet A is fully formed and all that remains is the docking of the N and C termini. We observe that in BF trajectories the final state may display single-strand crossovers which do not appreciably change the native contact map (Fig. S1). In particular, in the active, metastable A1AT structure, strands 3 and 4C cross over each other and the direction of this cross-over is reversed in the BF trajectories, see Fig. S1: however, this single strand cross-over does not alter the native contact map.

Standard MD simulations were started from representative frames taken from the BF trajectories. The structure was solvated with TIP3P water (35) and sodium and chloride ions were added to neutralize the system. Simulations were carried out at 300 K using periodic boundary conditions and the CHARMM36 force field (36). Electrostatics were calculated using the Particle Mesh Ewald method (37). Titratable groups were assigned protonation states for pH 7.0. The energy minimized system was heated to 300 K in 10 degree increments followed by a production run of 200 ns.

**Change-Point Analysis.** To identify the main transition points of the WT folding trajectory we adopted the multivariate change-point analysis introduced in reference (38) and implemented in the SIMPLE algorithm, developed by DE Shaw Research. This method is based on a statistical analysis of multi-dimensional time series extracted from atomistic trajectories. In particular, from one of our WT atomistic folding trajectories obtained using the BF approach, we extracted the evolution of the distances $r_{ij}$ between all alpha-carbons $i$ and $j$ which are in contact in the native state. For each of such pairs of residues we computed the time series of the function of $d_{ij}$ which interpolates smoothly between 0 and 1:

$$d_{ij} = \frac{1 - (r_{ij}/r_0)^6}{1 - (r_{ij}/r_0)^{10}} \qquad (3)$$

Where $r_0$=0.75 nm is a fixed reference distance. The sensitivity parameter λ of the change-point analysis (see reference (38)) was set to 7500 in order to yield four conformational change points.

**Folding Pathway Representation.** The density plots are obtained by projecting the full ensemble of LBTs onto the plane defined by the RMSD to the native structure and the fraction of native contacts Q. For WT, the native structure is PDB: 1QLP (31) and the native structures for the mutants are 1QLP mutated using the VMD mutator function (33). The high density regions reveal the existence of long-lived intermediates. For the sake of clarity, the density plots were smoothed using a Gaussian interpolation scheme. The arrows in the plots represent the preferential directions of the folding pathway and their widths are proportional to the number of trajectories that proceed from a given long-lived intermediate to another intermediate. The representative conformations shown in the plots are derived as described in next section, conformation harvesting.

**Conformation Harvesting.** We adopted the following criteria to harvest a number of representative conformations from the LBTs for each of the long-lived conformational ensembles.

Firstly, we selected two coordinates, i.e. the fraction of native contacts Q and the RMSD to the native structure, R. Given the histogram *P(Q, R)*, the conformations are harvested as follows:

1. We identified the minimum values of $G_B(Q, R) = -k_B T \log(P(Q, R))$ within the long-lived basins where $k_B$ is the Boltzmann constant and simulation temperature T = 300K;

2. We used $G_B$ as a rough estimate of the free energy landscape. Representative structures were randomly chosen from the conformations which deviate less than three $k_B T$ units from the corresponding minimum of $G_B(Q, R)$, while the cartoon representation is randomly chosen from those conformations which deviate at most one $k_B T$ unit from the minimum.

3. For each minimum, 25 conformations were selected in this way.

**Fraction of Native Contacts Distribution.** The frequency histograms of values of Q visited by the LBTs were obtained according to the following procedure. Let us consider a dataset composed of trajectories obtained from MD or BF simulations. For the whole collection trajectories we are interested in computing the fraction of native contacts, defined as

$$Q_0 = \sum_{i<j+3}^{N_R} C_{ij}^{\text{binary},0} \qquad Q(t) = \frac{1}{Q_0} \sum_{i<j+3}^{N_R} C_{ij}^{\text{binary}}(t) \tag{4}$$

where $N_R$ is the number of residues, $C^{\text{binary},0}$ is the binary contact map of the native structure, $C^{\text{binary}}(t)$ is the binary contact map at time *t* and $Q_0$ is the total number of native contacts in the active crystal structure. The probability of finding, among all the trajectories, a particular conformation identified by a fraction of native contacts is provided by

$$p(Q) = \frac{\sum_{T=1}^{M} \left[\sum_{\bar{Q}} \delta(Q - \bar{Q})\right]_T}{\int dQ \sum_{T=1}^{M} \left[\sum_{\bar{Q}} \delta(Q - \bar{Q})\right]_T} \tag{5}$$

where the sum is extended over all M trajectories. In practice, the Dirac delta is smeared and computed by choosing a fixed bin width and counting how many conformations fall within *Q* and *Q+dQ* for each trajectory in the set. This discrete distribution clearly shows which regions of native contacts are highly populated and which are rarely populated, thus giving an explicit indication of possible barriers to folding and the ability of a set of trajectories to reach, for example, the native state.

**Statistical analysis of Native Contacts Distributions.** To assess the statistical significance of differences between the histograms of native contacts shown in Figure 4, we performed a two-sample Kolmogorov-Smirnov (KS) test. In this test we compared the set of Q values observed in the folding/misfolding trajectories of the different A1AT variants (e.g., WT vs. Z, Z vs. S, and so on). Because the number of LBTs differs between variants, the number of Q values used to construct the histograms also changes from variant to variant. In order to determine if the number of LBTs (and thus Q values) for each variant is statistically robust, we bootstrapped from every population a number of values of Q corresponding to 1/2 of the population in the smallest dataset which contains 29 LBTs. More specifically, we used 18192 values of Q and ran 10 bootstraps for every KS test. Our null hypothesis for the KS test assumed that the Q values obtained by simulating the folding of different A1AT variants were sampled from the same distribution. In all the KS tests, the null hypothesis was rejected with a p-value $p < 2 \times 10^{-16}$. Based on these results we conclude that all of the histograms describe statistically independent distributions. Furthermore, since the bootstrapping involved a fraction of the smallest dataset, we can conclude that the number of simulations performed for all of the variants is sufficient to provide statistically representative samples of the corresponding Q distributions. Finally, to further enforce the robustness of our results, we computed the KS test both using R (39) and SciPy (40), obtaining comparable outcomes.

As a final remark, we stress that despite the fact that the number of successful folding events is small for all A1AT variants, our main result is that the Z (Glu342Lys) mutation induces a shift of the peak of the frequency distribution to a lower (i.e. less native like) value of Q. Again, this shift is statistically significant.

**Path Similarity Analysis:** The similarity parameter measures the consistency of the temporal succession in which native contacts are formed in two given pathways. The parameter takes on values ranging from 0 for no similarity, to 1 when all native contacts form in exactly the same succession for the two trajectories.

To compute this quantity, we define a matrix *M* which describes the order of native contact formation between atoms. Given as the time of formation of the i-th native contact in the k-th trajectory, the matrix element of the k-th path is defined as:

$$M_{ij}(k) = \begin{cases} 1 & \text{if } t_{ik} < t_{jk} \\ 0 & \text{if } t_{ik} > t_{jk} \\ \frac{1}{2} & \text{if } t_{ik} = t_{jk} \end{cases} \quad (6)$$

The similarity parameter is thus defined as

$$s(k, k') = \frac{1}{N_c(N_c - 1)} \sum_{i \neq j} \delta\left(M_{ij}(k) - M_{ij}(k')\right) \quad (7)$$

where $N_C$ is the total number of native contacts. To provide a robust indication of the degree of heterogeneity of multiple trajectories, we can also compute the distribution of *s*, path similarity, over all possible pairs of trajectories, defined as

$$p(s) = \sum_{k<k'} \delta\left(s - s(k, k')\right) \quad (8)$$

The path similarity calculation allows us to quantitatively compare trajectories for a single A1AT variant, e.g., to determine similarities and differences between WT folding trajectories, and to determine whether and when the folding of A1AT mutants diverges from WT-like folding.

**Solvent Accessible Surface Area (SASA) graphs.** SASA graphs have been obtained by using the Shrake-Rupley function implemented in MDTraj 1.9.0 (41) using a probe radius of 0.144 nm and 100 sphere points.

**Results and Discussion**

**The BF Approach.** Using the BF algorithm (23) it is feasible to use standard computer clusters to generate many folding and misfolding events for proteins as large as serpins, using state-of-the-art all-atom force fields. This algorithm identifies the most realistic reaction pathways within an ensemble of uncorrelated trial trajectories, generated by the so-called ratchet-and-pawl MD (rMD) (24, 25). In rMD, a biasing force is introduced only when the chain tends to back-track towards the unfolded state and it is not applied when the chain spontaneously progresses towards the native state. In the BF approach a rigorously derived variational condition is applied to identify the trial rMD pathways which have the largest probability to occur in the absence of any bias. In the following, we shall refer to these paths as Least Biased Trajectories (LBTs).

The accuracy of *all* variational approaches critically depends on the quality of the model subspace, i.e. the set of possible solutions, within which the optimal approximate solution is identified through the variational condition. In the BF approach, the model subspace is defined by the trial folding trajectories generated using the rMD protocol. The BF method thus provides the relative statistical weights of the trial trajectories in our variational space, rather than absolute statistical weights for individual trajectories. Since a suboptimal choice of the model subspace may introduce systematic errors, variational approaches must be carefully benchmarked against exact methods and extensively validated against experimental data.

The BF method was benchmarked against the results of MD folding simulations, performed using the Anton supercomputer (23, 28), where both methods used the same all-atom force field. In particular, the folding mechanism and the precise order of native contact formation during the folding of an all-beta protein (WW domain Fip35) and of an all-alpha protein (villin headpiece subdomain) predicted by the BF method were found to be statistically indistinguishable from those obtained using conventional MD methods. The predictions of the BF approach have also been validated against experimental data. For example, the BF method was used to study the folding of two alpha proteins consisting of nearly 100 amino acids (IM7 and IM9). These two proteins have highly homologous native structures, but very different folding mechanisms; Im7 folds via a three state mechanism while Im9 folding is two state (42). The BF method correctly reproduced these experimental observations (43). Furthermore, in a very recent investigation (30) the BF approach has been interfaced with quantum electronic structure calculations to yield a direct prediction of the expected time-resolved near and far UV circular dichroism (CD) spectra in the folding of the proteins canine lysozyme and IM7. The BF results quantitatively agree with the experimental data, correctly predicting the existence of a folding intermediate and reproducing the difference between the intermediate and native state near UV spectra, which reports on the local environment of aromatic residues.

An early version of the BF algorithm (called dominant reaction pathway or DRP) was also successfully applied to sample the conformational transition leading to serpin latency (32). These BF simulations predicted the atomistic structure of an intermediate in the active to latent transition for the serpin plasminogen activator inhibitor (PAI-1) and provided an atomistically detailed picture which explained why binding a specific small molecule accelerates the PAI-1 latency transition. For both PAI-1 and A1AT the BF simulations reproduced the experimentally observed propensity of specific point mutations to accelerate or retard the latency transition. This study also delineated differences between the PAI-1 latency transition and that of A1AT to help explain why PAI-1 easily accesses the latent state while A1AT does not.

In order to further explore serpin conformational changes and to validate the BF simulation results, we begin the present study by simulating and analyzing the folding of the WT A1AT. We compare the results of the simulations to available experimental results and provide new, *testable* atomistically resolved data on how WT A1AT folds. These WT simulations also provide the reference for comparing folding and misfolding pathways. We therefore proceed to simulate and analyze the folding of disease associated A1AT variants and suppressor mutants.

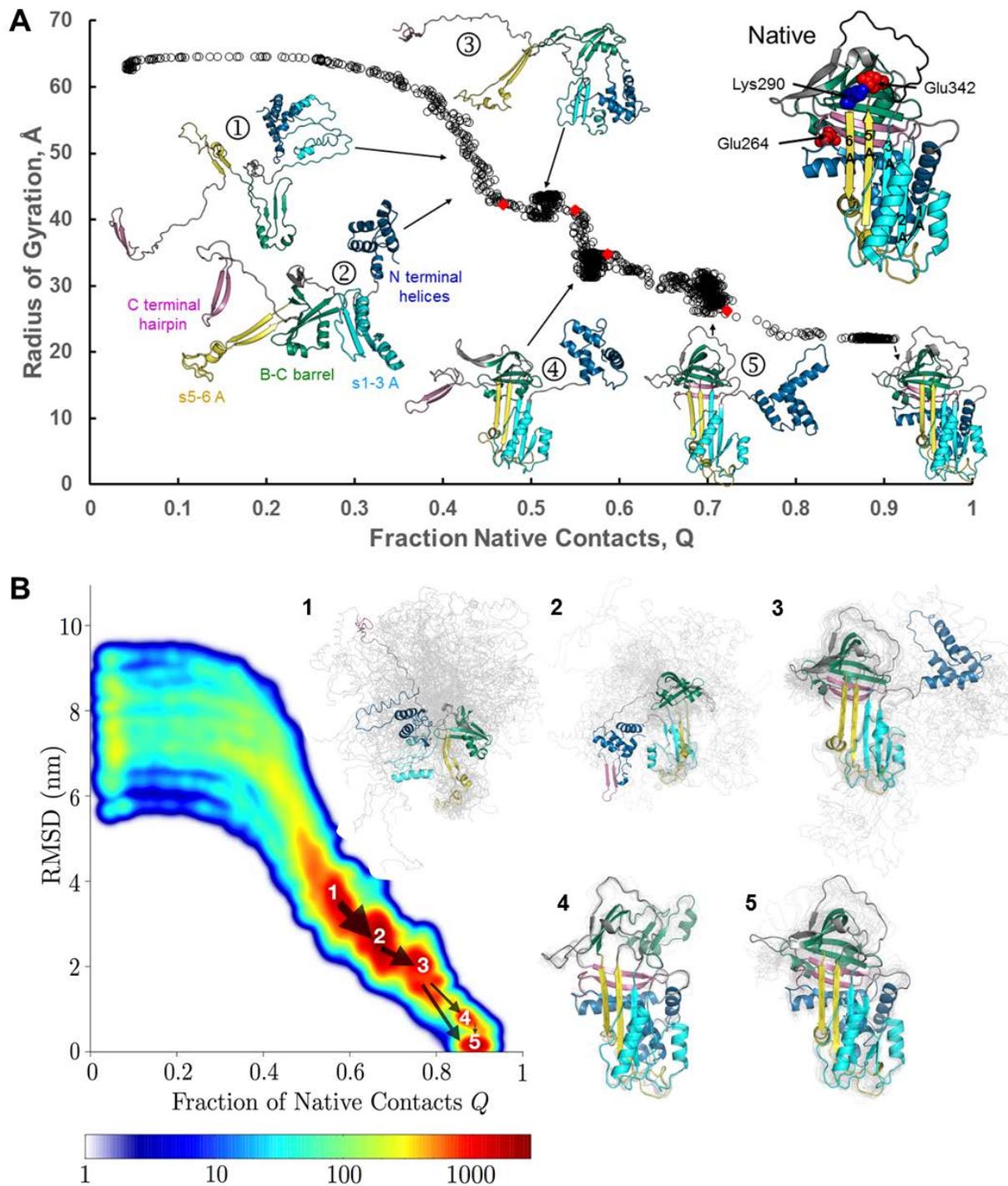

**Figure 2.** WT A1AT folding. A) Changes in the WT radius of gyration, $R_g$, as a function of native contacts formed along a typical successful WT folding trajectory. Change-points (38) are indicated by red diamonds. The A1AT structure colored by foldon and indicating the locations of the Z and S mutations and Lys290 is shown on the upper right. B) Kinetic free energy landscape generated from all frames in all WT LBTs which reach the native state, colored according to population density from blue (low populations) to red (high populations). The arrows indicate fluxes between the labeled local minima

and arrow widths are proportional to the flux. Representative conformational ensembles from each of the labeled minima, harvested as described in the Methods, are shown to the right in gray with a single representative conformation colored as in A.

**WT A1AT Folding Pathways.** For all of the A1AT variants, folding starts with the independent formation of local structures (Fig. 2A, stage 1) that we refer to as foldons, following the usage of Wolynes, Englander and coworkers (44, 45). In the majority of successful WT trajectories, foldons dock in a well-defined order, as determined from visual inspection, from the plot of the radius of gyration versus fraction of native contacts and from an automated statistical analysis that identifies the most relevant change-points (38) (Fig. 2A).

In WT LBTs, initial foldon formation is followed by the formation of native interactions between residues at the top of strands 5 and 6A (s5/6A) in the nascent sheet A and the nascent B-C barrel formed by parts of strands 3C, 4C and 1 to 3B (Fig. 2A, stage 2). In particular, a hydrogen bond is formed between Glu342 and Thr203 and van-der-Waals interactions are established between Pro289 and Met residues 220 and 221. To assess the role of rMD forces in maintaining these interactions, the structures corresponding to stages 1 (foldon formation) and 2 (early inter-foldon interactions) in Figure 2A were subjected to 200 ns of conventional MD simulations in explicit solvent. In these MD simulations, the C-terminal β–hairpin formed by strands 4/5B is not stable consistent with the experimental finding that the isolated peptide composed of the 36 C-terminal A1AT residues lacks stable structure (46). In contrast, as shown in Figure 3, the three N-terminal foldons and the network of interactions between the nascent B-C barrel and s5/6A are stable on the time scale of these conventional MD simulations even in the absence of any additional stabilizing forces.

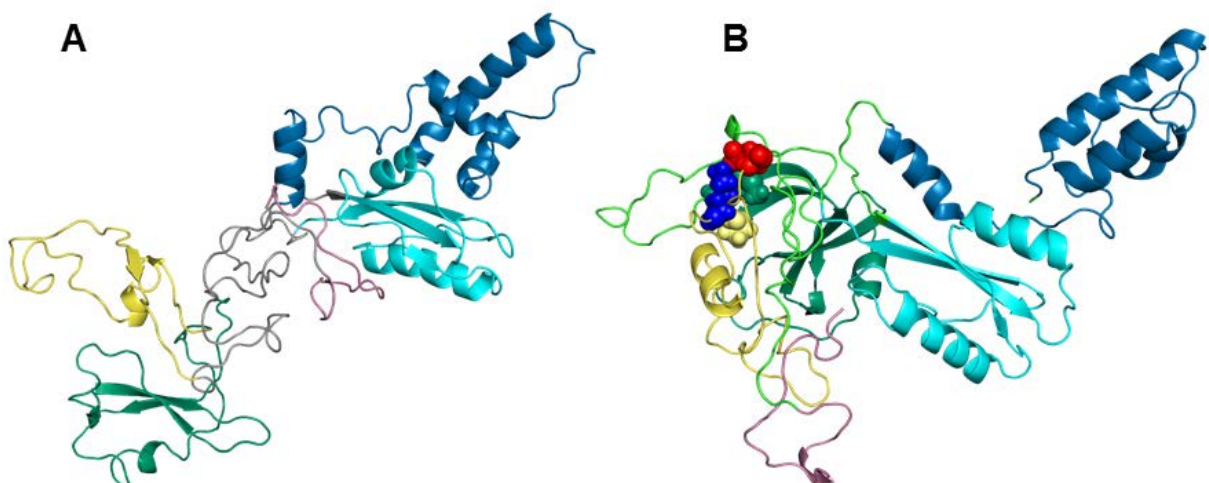

**Figure 3.** A) Stability of A1AT foldons. The structure from stage 1 in Figure 2A after 200 ns of conventional MD simulations in explicit solvent. The three foldons in the N-terminal two-thirds of A1AT are stable on this time scale. B) Stability of the contacts between s5/6A and the B-C barrel. The structure from stage 2 in Figure 2A after 200 ns of conventional MD simulations in explicit solvent. Glu342, red:

Lys290, blue: Pro289, yellow: Met220 and Met221, green: Thr 203, grey. The ribbon structure is colored by foldon as described in Figure 2A.

In most of the WT trajectories that reach the native conformation (see Movie S1), non-native interactions between the loop at the top of strand 3A and β-strands 2 and 3B prevent correct positioning of the B-C barrel leading to the first barrier (Fig. 2A, stage 3). Disruption of this steric hindrance allows for the correct positioning of the B-C barrel and docking of s5/6A to s1-3A thus completing β-sheet A formation (Fig. 2A, stage 4). At this stage, the C-terminal β–hairpin (s4/5B) and the N-terminal helices are solvent exposed and free to move on flexible linkers. Folding completes when the C-terminal hairpin docks to strands 1 to 3B (Fig. 2A, stage 5) followed by packing of the N-terminal helices and s6B on the back of the β sheets.

While the results from one successful WT folding trajectory are informative, further insight into A1AT folding can be gained by studying patterns that emerge from *all* of the successful LBTs. To this end, a density plot was generated by projecting all frames from all successful WT simulations onto the plane defined by root mean squared distance (RMSD) from the native X-ray crystal structure (PDB: 1QLP (31)) and fraction native contacts, Q (Fig. 2B). In this projection, metastable states (local energy minima) are identified as densely populated regions (red) separated by sparsely populated regions corresponding to barriers. Due to the action of the rMD force, the barriers in this representation can only provide lower bounds on the true free energy barriers. This representation can therefore be thought of as a kinetic free energy landscape. Also shown are representative conformational ensembles (gray) harvested from each basin with a representative structure shown in color.

As expected, the metastable states in the WT density plot correlate with the most highly populated intermediates along the single LBT (Fig. 2A), with the addition of a new near native metastable state (basin 4) associated with a minor folding pathway in which consolidation of the B-C barrel is the last folding step. Local minima corresponding to the progressive docking of the initially formed foldons are clearly evident, identifying the major successful folding pathway. Closer inspection of the major basins and the fluxes between them reveals that for the major pathway the order of folding events is: initial non-native interactions between the foldons, the completion of sheet A while both N and C termini remain exposed, docking of the C terminal β hairpin and finally docking of the N terminal helices, in agreement with the single representative trajectory in Figure 2A.

Our finding that completion of sheet A precedes C-terminal hairpin incorporation into sheet B agrees with fragment complementation studies, where docking of a fragment containing the C-terminal hairpin with a larger N-terminal A1AT fragment requires the presence of s5A (11) and thus, presumably, completion of sheet A. Similarly, in kinetic refolding experiments monitored by oxidative labeling of sidechains or hydrogen/deuterium exchange of backbone amides detected by mass spectrometry (MS), s5A is one of the last regions to acquire native-like protection (12, 13). These same MS based studies found that s4B remains solvent exposed until late in the folding process, consistent with the late packing of the C-terminal hairpin (s4/5B) seen in our simulations.

While A1AT has only a single Cys residue and no disulfide bonds, folding studies of two serpins containing disulfide bonds, ovalbumin and antithrombin III, show that N-terminal disulfide bond formation and rearrangements occur after the C-terminus packs (47, 48). These experimental results imply that the N-terminus is involved in the final stage of serpin folding. The large conformational transition performed by the N-terminal foldon in the last stage of WT A1AT folding suggests that the last steps in A1AT, ovalbumin and antithrombin III folding are similar and provides a *testable*, atomistically detailed explanation for the experimental results. Recently, the folding of A1AT was investigated using a coarse grained 1 bead per residue Go model (49). Detailed comparisons with the present study are difficult due to the different representations of the polypeptide chain, differences in the force fields used and the high temperature used for the Go model equilibrium simulations. Nonetheless, while the details differ, we do note broad agreement between the two studies in that formation of the B-C barrel region precedes full formation of the domain containing sheet A and the N-terminal helices.

The BF WT A1AT folding simulations are in good agreement with existing serpin folding data and set the stage for generating detailed models of mutation induced serpin misfolding.

**Misfolding of A1AT Disease-Associated Variants.** BF folding simulations for the S (Glu264Val) and Z (Glu342Lys) mutants suggest that these two A1AT variants misfold differently. To effectively analyze the differences between WT, S and Z folding and misfolding for these non-equilibrium BF folding simulations we focused on the observation that mutants with a higher misfolding propensity are less likely to reach conformations in which almost all of the native contacts have been formed. To quantify this measure for each A1AT variant, we used all of the frames from all of the LBTs generated for each A1AT variant to construct a histogram describing how often each variant visits conformations with a given fraction of native contacts, Q (Fig. 4A). Comparisons between such histograms for WT and Z simulations correctly predict that the pathological Z mutant is significantly less able to populate structures close to the native state, and the statistical significance of this difference is confirmed by a Kolmogorov-Smirnov test. In fact, none of the Z mutant LBTs successfully reached the fully folded state. In contrast, the Q histograms for S and WT trajectories are similar, consistent with the observation that the S mutation has only mild effects on A1AT folding (1) (Fig. 4A).

The Z mutation disrupts a network of electrostatic and hydrogen-bonding interactions at the top of β-sheet A (Fig. 4B). Experimentally, Z misfolding can be rescued by mutating Lys290 to Glu (50) leading to a reversed version of the original salt bridge (Lys290Glu/Glu342Lys or Glu/Lys) and reducing the probability of steric clashes. Consistent with experiment, our Glu/Lys simulations show a WT like distribution of native contacts (Fig. 4A).

One way to test the relative roles of electrostatics and sterics in Z misfolding is to place Glu at both 342 and 290 (Lys290Glu/Glu342 or Glu/Glu) thus preserving the electrostatic repulsion present in Z but reducing the side chain length. Consistent with cellular studies, where Z and Glu/Glu shows approximately 20% and 75% secretion efficiency, respectively, relative to WT A1AT (51), the distribution of native contacts in the Glu/Glu simulations resembles that of WT (Fig. 4A, Movie S2). The fact that both experiments and computations indicate that Glu/Glu can fold with reasonable efficiency suggests that the 290/342 salt bridge may not be essential for A1AT folding. To further test this hypothesis we determined whether replacing Lys290 with Ser could rescue Z (K290S/E342K or Ser/Lys) simply by alleviating steric clashes. The resulting LBTs are more WT-like than Z-like (Fig. 4A).

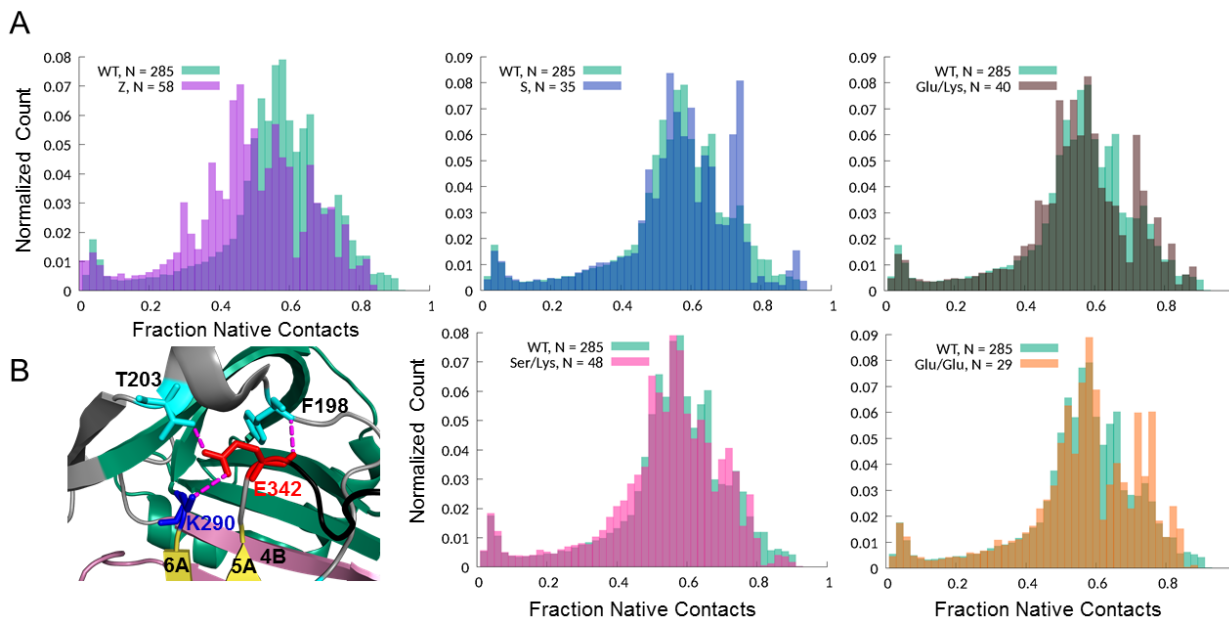

**Figure 4.** A) Comparisons of the fraction of native contacts, Q, formed during folding and misfolding LBTs. For the indicated variants with N total LBTs, histograms were calculated using all frames from all N trajectories. The data for Z (E342K, N=58) and S (E264V, N=35) are in purple and blue respectively. Suppressor variants with mutations at residues 290 and/or 342 are Glu/Glu (K290E/E342, N=25), Glu/Lys (K290E/E342K, N=40) and Ser/Lys (K290S/E342K, N=48) in orange, gray and pink, respectively. In all of the histograms, the WT (N=285) data are in green, and darker colored regions represent the overlap between WT and mutant distributions. B) Close up view of the network of electrostatic and hydrogen-bonding interactions formed by Glu342 in WT A1AT.

Further insight into the effects of mutations on intermediates populated during A1AT folding/misfolding can be gained by comparing the kinetic free energy landscapes of WT and the various mutants (Figs. 5, S2 and S3). The Z mutant energy landscape is markedly different from WT. In addition to the fact that none of the Z trajectories achieve more than 80% of the total native contacts, all of the major free energy basins in the Z landscape are populated by structures that are less folded than the (roughly) corresponding WT intermediates. This result suggests that the folding of Z diverges from that of WT early in the folding pathway. Even in the most native like structures attained by Z, sheet B fails to fully form, leaving strands 4 and 5B solvent exposed (basin 5 in Fig. 5). However, few trajectories reach this state, and most are trapped in more unfolded conformations in which both sheets A and B fail to fully form. It is known that Z polymerizes from a partially folded conformation (4, 52), although the nature of this species, like the structure of the resulting polymers, is a matter of debate. The heterogenous mixture of partially folded Z conformations generated by the BF folding simulations make it difficult to identify a single candidate polymerization prone species.

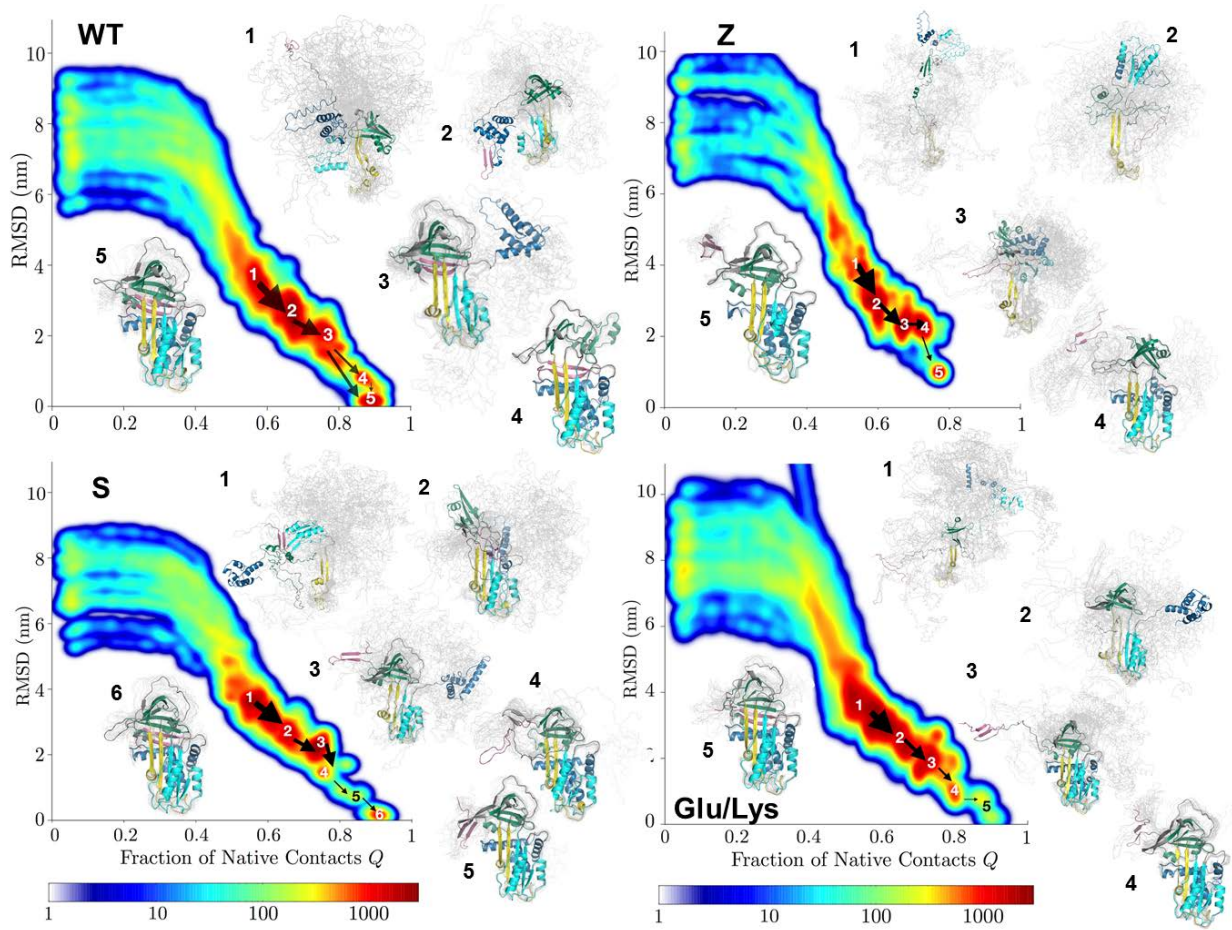

**Figure 5**. Kinetic free energy landscapes for WT A1AT and the Z, S, and Glu/Lys variants. Also shown are ensembles of representative structures from the identified basins generated as described in Materials and Methods and Figure 2B. Landscapes and ensembles for the Ser/Lys and Glu/Glu mutants are shown in Figures S2 and S3.

Since the majority of the trajectories result in Z conformations with incomplete β sheets, we selected one such representative structure and subjected it to conventional MD simulations in explicit solvent in order to allow secondary structure to relax in the absence of rMD forces. After 100 ns sheet A and half of sheet B as well as regions of sheet C are substantially disordered (Fig. S4). Nevertheless, enough secondary structure is retained to give a calculated CD spectrum (53) that is more characteristic of a folded protein than an unstructured coil. The structure is also relatively compact, with a radius of gyration of 24.9 Å$^2$ only 12% larger than that of native A1AT (21.9 Å$^2$). Ekeowa *et al*. used ion mobility mass spectrometry to measure the collisional cross section (CCS) of a polymerigenic species of WT A1AT that is populated upon mild heating and has been proposed as a model for the polymerization prone form of Z (54). They found that the CCS of the polymerigenic species is 118±17% that of the folded monomer. The CCS of our misfolded Z structure after 100 ns of MD is 123% that of folded WT A1AT (CCS calculated from structures using the program IMPACT (55)). All of these results suggest that Z misfolds to a state that, in terms of compactness and secondary structure, has substantial native-like character, consistent with arguments that Z polymerizes from a state that is at least somewhat native-like rather than globally unfolded (52).

In contrast to Z, the energy landscape of the mildly deficient S mutant resembles that of WT (Fig. 5), and the major intermediates populated during folding are more structured than those of Z. The same is true for the Glu/Lys rescue mutant. Interestingly, in both of these variants, the order of the C and N termini docking is the opposite of WT. In fact in S and Glu/Lys, Ser/Lys, and Glu/Glu, the N terminal helices dock first, followed by packing of the C terminal hairpin (Figs. 5, S2 and S3). In the case of S, this delayed C-terminal packing is likely explained by the fact that the Glu264Val mutation eliminates a salt bridge between Glu264 in helix G and Lys387 in the C terminal hairpin (56). This electrostatic interaction would favor docking of the C terminus. Why C-terminal packing is delayed is less clear for Glu/Lys and the other rescue mutants, but the relative orientations of the B-C barrel and sheet A are subtly different from that observed in WT (Fig, 5).

In S and WT trajectories misfolding occurs late, resulting from premature docking of the N-terminal helices, which leaves the C-terminus solvent exposed (Fig. 5, Movie S3). Premature docking of the N-terminal helices to the rest of A1AT is essentially irreversible in the BF simulations due to the number of native contacts gained in these interactions. However, in experiments this misfolding may be reversible, and in cells interactions between the lectin chaperones and the N-terminal glycans at Asn residues 46 and 83 could help protect against this late misfolding. 100ns of standard MD simulations of misfolded S with the C terminus exposed show that it remains structured on this time scale (Fig. S5). In contrast misfolded Z loses ordered structure during simulations on this same time scale (Fig. S4). This difference in stability mirrors the relative pathogenicity of these two mutations (1, 52).

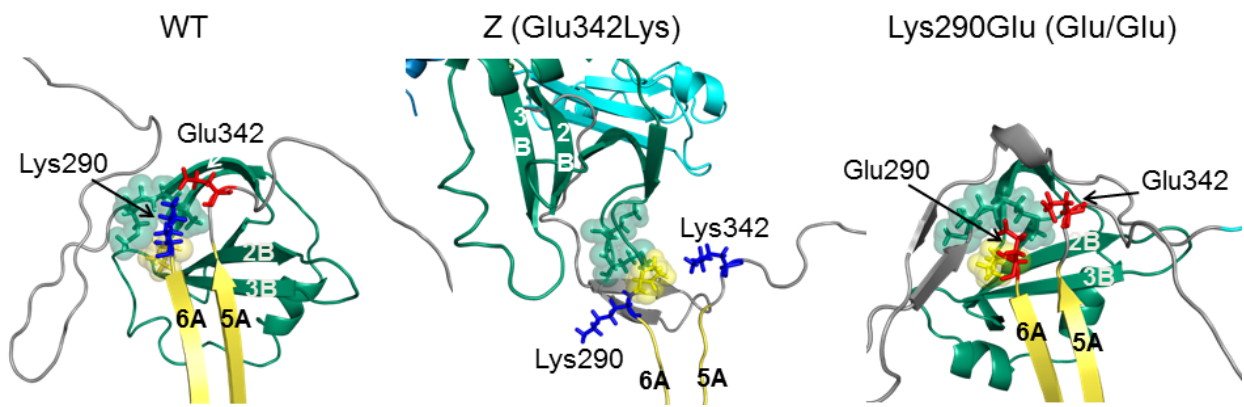

**Figure 6.** The divergence of WT and Z folding with close up views of the region containing residues 290 and 342 (E: red and K: blue) and P289 (yellow) with T203, M220 and M221 in green. Glu/Glu is also shown. Structures are representative snapshots taken from basin 2 (Figs. 2 and 5) of the BF simulations.

In striking contrast, Z and WT folding diverge much earlier (Movie S4). As discussed above, the Z mutation replaces a conserved salt bridge between Lys290 and Glu342 (Fig. 6) with a charge-charge repulsion. This mutation appears to disrupt early native and non-native interactions between β-strands 5/6A and the B-C barrel. Electrostatic repulsion and steric hindrance, due to the length of the two Lys sidechains (Lys290/Glu342Lys), increase the probability that these residues assume a non-native spatial orientation. In contrast to Z, Glu/Glu achieves WT like docking between s5/6A and the B-C barrel despite the electrostatic repulsion between the two Glu residues. This supports the hypothesis that sterics, in addition to electrostatics, play a key role in Z misfolding.

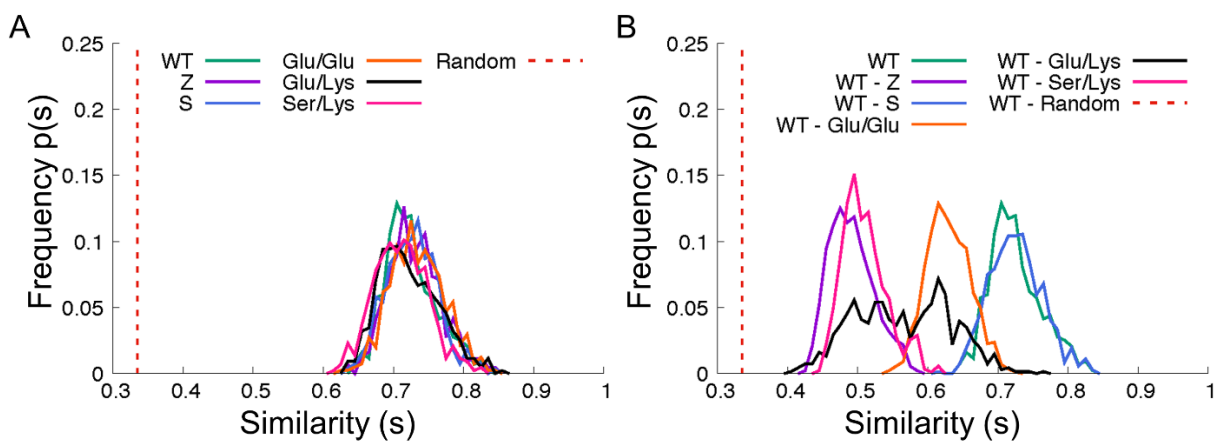

**Figure 7.** Distributions of path similarities for WT and mutants (see Methods). The vertical dotted line indicates the average similarity score for two random sequences of native contact formation. Self-similarities are shown in A. Differences between the WT paths and the various mutant paths are shown in B. The coloring is the same as in Figure 4A.

**Investigating the Order of Contact Formation.** While the various rescue mutants restore the correct docking of s5/6A to the B-C barrel, it can be seen from the ensembles in Figures 5, S2 and S3 that they do not necessarily restore the full WT folding pathway. To investigate this more quantitatively, we performed a statistical analysis based on the distribution of path similarity s introduced in (57) and defined in the Methods. This parameter is equal to 1 when native contacts in two reactive trajectories are formed in exactly the same order, whereas s=0 when the order is entirely different. For completely random sequences of native contact formation the similarity distribution is sharply peaked around 0.3. Path similarities calculated between LBTs for the same variant (self-similarity) or between trajectories of different variants (cross similarity) are shown in Figure 7. For any given A1AT variant the self-similarity distribution obtained from all of the LBTs peaks around 0.6 (Fig. 7A). In contrast, comparisons between different variants show that Z LBTs differ significantly from WT (peak s~0.4), while S and WT pathways are similar (peak s~0.6) (Fig. 7B).

Path similarity analyses also indicate that while mutations at Lys290 may suppress Z misfolding, the paths are very different for different combinations of residues at positions 290 and 342. Folding of the Glu/Glu variant, where Lys290 is mutated to Glu in the WT background providing an anionic charge repulsion between residues 290 and 342 compared to the cationic repulsion in the Z mutant, is most WT-like. The folding of the Ser/Lys variant diverges the most from WT, while the order in which contacts are formed in Glu/Lys - the charge-reversed salt-bridge in the Z background - span a large range from WT-like to quite different (Fig. 7B).

We note that the native state contact map, which defines the collective coordinate along which the biasing force acts, is not significantly affected by the mutations, as the structures of WT and the mutants are nearly indistinguishable apart from the mutated residue itself. This is unlikely to be due to the fact that we introduced the mutations computationally, as the crystal structure of the Z mutant was recently published, and its structure is essentially superimposable on that of WT (34). Thus, the collective coordinate along which the biasing force acts was essentially identical for all variants studied. This strongly suggests that the differences observed are due to the sterics and energetics of the mutations themselves, rather than to the biasing force.

These results, and the finding that in the BF folding simulations all of the salt-bridge variants are more likely to fold than Z, suggest that A1AT can fold using a number of alternative folding pathways. Moreover, they show that the Lys290-Glu342 salt-bridge, while not required for folding, is important for increasing the probability of the major folding mechanism observed in the WT A1AT BF folding simulations.

**Experimental Observables.** WT A1AT contains two Trp residues, 238 in strand 2B, and 194, C-terminal to strand 3A in the "breach" region at the top of sheet A. Studies of A1AT single Trp mutants, in the WT background, show that during equilibrium unfolding, Trp194 shows 2 distinct transitions corresponding to a native to intermediate transition followed by an intermediate to unfolded transition. In contrast, Trp238 shows a single broad transition (58). In addition, stopped flow studies of WT A1AT refolding monitored by Trp fluorescence emission reported at least three kinetic phases for A1AT WT refolding, a very fast, ~50 ms, phase; a slower, ~500 ms, phase and a very slow 100s of seconds long phase (9). Trp194 fluorescence may be altered both by changes in solvent accessibility and by quenching by Tyr244. Consistent with experimental observations, the Trp194-Tyr244 distance changes in 2 stages, one corresponding roughly to the formation of the foldons and the other occurring with the docking of strands 5 and 6A to the B-C barrel (Fig. 8A). For WT A1AT Trp194 burial occurs approximately in a single step during the completion of β-sheet A (Fig. 8A). Trp238 burial does not occur in a single step. Rather, it shows an initial large decrease in accessibility, followed by a gradual decrease and then a later sharp drop. This is qualitatively consistent with the observation that Trp238 shows a single broad transition during equilibrium unfolding. The multiple transitions in the local Trp environments in our simulations are consistent with the multiple kinetic phases seen when refolding is monitored by Trp emission (9). Comparison of WT with S and Z (Fig. 8B) and the rescue mutants (Fig. S6) show that Trp194 is buried earlier in WT than it is any of the mutants. This computational observation could potentially be tested using kinetic folding experiments monitored by Trp lifetimes or fast oxidative footprinting coupled to MS.

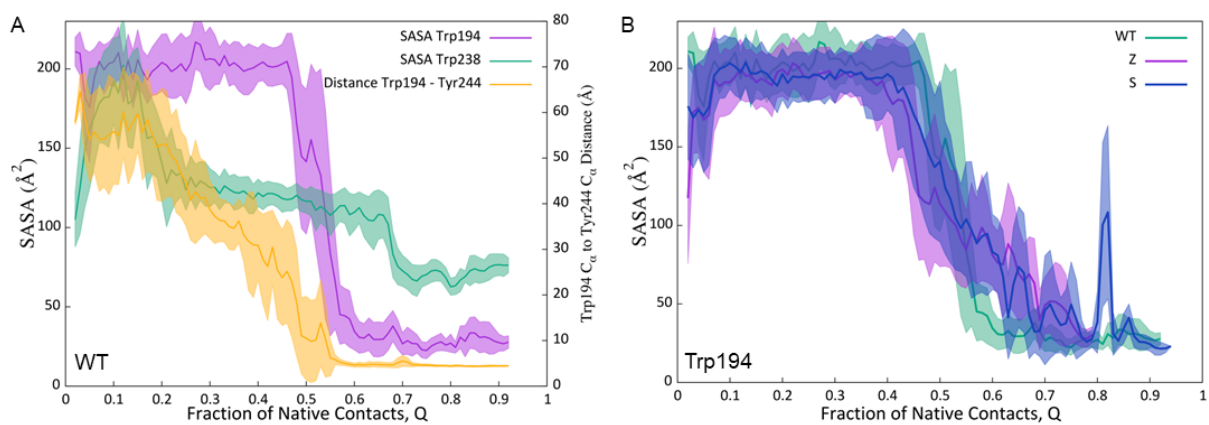

**Figure 8.** A) Evolution of solvent accessible surface areas (SASA) for Trp194 and Trp238 and the Trp194-Tyr244 distance during WT A1AT folding. Solid lines show the averages over all successful simulations while shaded areas show the standard deviation. B) Evolution of Tryp194 SASA during folding of WT and the S and Z mutants.

The A1AT BF folding simulations suggest that the biggest folding differences between WT and other variants involve the relative timing for packing the N-terminal helices and the C-terminal β hairpin. Perhaps the best way to monitor this difference is by looking at how the N-terminal helices/ β sheet A and the β sheet A/C-terminal hairpin distances change as A1AT variants fold. Kinetic refolding experiments monitored by Förster resonance energy transfer (FRET) have the potential to reveal such differences. A1AT contains a single Cys, Cys232 in s1B, and a number of amino acids in A1AT have been mutated to Cys without significantly impairing function or stability (59–62). In particular, both Ser47 in the N-terminal helical region and Ser313 in the loop from helix I to strand 5A have been mutated to Cys. Calculation of the 47 to 232 and 47 to 313 distances suggest that FRET pairs at these locations are likely to be sensitive to docking of the N-terminal helices (Fig. S7). Similarly, comparative kinetic refolding experiments using oxidative labeling of sidechains monitored by MS would provide a rigorous test of the BF predicted folding mechanisms since foldon docking should significantly alter side chain accessibility.

**Conclusion**

In conclusion, our calculations provide a coherent atomistic and physics-based picture of serpin folding and misfolding. For WT A1AT, we find that there is a major folding pathway that begins with the initial assembly of local structural units, followed by higher order associations, some of which involve non-native contacts. The pathway ends with the incorporation of the C-terminal β–hairpin followed by docking of the N-terminal helices. These findings are supported by existing experimental data (9–13, 47, 48, 50, 51, 58) and the detailed molecular mechanisms provided here are experimentally testable. The multi-dimensional free energy landscape for folding is complex and, as exemplified by the Z suppressor mutants, there are alternative ways to successfully fold to the native state. Our simulations of pathological and suppressor mutants also elucidate the mechanism of Z misfolding and make the prediction that misfolding occurs early in the folding process, a prediction that should be amenable to experimental testing.

The scheme presented in this work opens a way to investigate many important disease-associated processes, which occur over minutes or hours, using only ordinary medium-sized computer clusters of the type available to most computational laboratories.

**Acknowledgements:** This work was funded by the Alpha-1 Foundation (PLW, PF) and NIH grant GM094848 (AG). We thank DE Shaw Research for providing the code for performing change-point analysis. We also thank several anonymous referees for valuable suggestions.

**Author Contributions:** F.W. designed, performed and analyzed the simulations and wrote the paper; S.O. analyzed the simulations and, as required, developed new analysis methods, and wrote the paper; A.I. & G.S. analyzed the simulations and, as required, developed new analysis methods; S.a B. designed and wrote the software to perform BF simulations; P.F. designed and analyzed simulations, developed new analysis methods as necessary and wrote the paper; A.G. & P.L.W. designed and analyzed simulations and wrote the paper.

**Supplementary Information**

**Figure S1**. Strand crossover in the B-C barrel

**Figures S2 & S3**. Kinetic free energy landscapes and structures for the Glu/Glu and Ser/Lys variants, respectively.

**Figures S4 & S5.** Structures before and after 200 ns of conventional MD simulations for misfolded Z and S, respectively.

**Figure S6.** Trp SASA data for all of the A1AT variants.

**Figure S7.** Evolution of Ser47 to Cys 232 and Ser47 to Ser313 C-alpha distances for all WT BF folding simulations.

**Movie S1:** BF simulation of successful WT A1AT folding.

**Movie S2:** BF simulation of the successful folding of the Glu/Glu (Lys290Glu/Glu342) A1AT variant.

**Movie S3:** BF simulation of S (Glu264Val) A1AT misfolding

**Movie S4:** BF simulation of Z (Glu342Lys) A1AT misfolding

All-Atom Simulations Reveal How Single-Point Mutations Promote Serpin Misfolding

Fang Wang, Simone Orioli, Alan Ianeselli, Giovanni Spagnolli, Silvio a Beccara, Anne Gershenson, Pietro Faccioli, and Patrick L. Wintrode

**Supporting Materials and Methods**

**Force Fields.**

MD and rMD simulations were performed using the AMBER ff99SB force field (1), within the Generalized Born scheme implemented in GROMACS 4.5.2 (2). In general, the use of an implicit solvent model in protein folding MD simulations is considered a rather strong approximation. However, we note that previous applications of the BF approach and closely related variants used the same implicit solvent force field and produced results which compared well with those from MD simulations in explicit solvent (3, 4) and with experiments (5). Trial trajectories were generated by running 30 ns of rMD simulations at room temperature, 300 K, with an integration time of 1 fs. The parameter $k_R$ in Equation 1, setting the strength of the biasing force, was 0.02 kJ/mol.

**Simulation Details**

For all A1AT mutants (5), single point mutations were introduced using the mutator function in VMD (6). After we began our simulations, an X-ray crystal structure of the Z mutant was published, PDB: 5IO1 (7). There are no significant differences between the WT and Z active, native structures. A structural alignment of our target Z structure with the published structure shows a root mean square deviation (RMSD) of 0.6 Å and the native contact maps are identical.

For simulations of WT folding we generated 12 independent fully denatured conformations by thermal unfolding at 1600 K. From each such condition we produced 12 trial rMD trajectories and extracted the Least Biased Trajectory (LBT) using the minimum bias condition, i.e. selecting the path with the least BF functional, defined in Eqn. 2. Typically, for such a number of trial trajectories the requirements for convergence described in Ref. (8) are satisfied. For Z, S, and other A1AT variants, we also began each BF simulation with 12 independent unfolded conformations. Twelve LBTs of A1AT folding from 12 independent unfolded states can be generated in roughly 1 week using 32 cores on a standard computing cluster. This procedure was repeated multiple times yielding the total number of LBTs for each A1AT variant given in Figure 4A.

Standard MD simulations were started from representative frames taken from the BF trajectories. The structure was solvated with TIP3P water (9) and sodium and chloride ions were added to neutralize the system. Simulations were carried out at 300 K using periodic boundary conditions and the CHARMM36 force field (10). Electrostatics were calculated using the Particle Mesh Ewald method (11). Titratable groups were assigned protonation states for pH 7.0. The energy minimized system was heated to 300 K in 10 degree increments followed by a production run of 200 ns.

**Change-Point Analysis.** To identify the main transition points of the WT folding trajectory we adopted the multivariate change-point analysis introduced in reference (12) and implemented in the SIMPLE algorithm, developed by DE Shaw Research. This method is based on a statistical analysis of multi-dimensional time series extracted from atomistic trajectories. In particular, from one of our WT atomistic folding trajectories obtained using the BF approach, we extracted the evolution of the distances $r_{ij}$ between all alpha-carbons $i$ and $j$ which are in contact in the native state. For each of such pairs of residues we computed the time series of the function of $d_{ij}$ which interpolates smoothly between 0 and 1:

$$d_{ij} = \frac{1 - (r_{ij}/r_0)^6}{1 - (r_{ij}/r_0)^{10}} \qquad (3)$$

Where $r_0$=0.75 nm is a fixed reference distance. The sensitivity parameter $\lambda$ of the change-point analysis (see reference (12)) was set to 7500 in order to yield four conformational change points.

**Folding Pathway Representation.** The density plots are obtained by projecting the full ensemble of LBTs onto the plane defined by the RMSD to the native structure and the fraction of native contacts Q. For WT, the native structure is PDB: 1QLP (13) and the native structures for the mutants are 1QLP mutated using the VMD mutator function (6). The high density regions reveal the existence of long-lived intermediates. For the sake of clarity, the density plots were smoothed using a Gaussian interpolation scheme. The arrows in the plots represent the preferential directions of the folding pathway and their widths are proportional to the number of trajectories that proceed from a given long-lived intermediate to another intermediate. The representative conformations shown in the plots are derived as described in next section, conformation harvesting.

**Conformation Harvesting.** We adopted the following criteria to harvest a number of representative conformations from the LBTs for each of the long-lived conformational ensembles. Firstly, we selected two coordinates, i.e. the fraction of native contacts Q and the RMSD to the native structure, R. Given the histogram $P(Q, R)$, the conformations are harvested as follows:

1. We identified the minimum values of $G_B(Q, R) = -k_BT \log(P(Q, R))$ within the long-lived basins where $k_B$ is the Boltzmann constant and simulation temperature T = 300K;

2. We used $G_B$ as a rough estimate of the free energy landscape. Representative structures were randomly chosen from the conformations which deviate less than three $k_BT$ units from the corresponding minimum of $G_B(Q, R)$, while the cartoon representation is randomly chosen from those conformations which deviate at most one $k_BT$ unit from the minimum.

3. For each minimum, 25 conformations were selected in this way.

**Fraction of Native Contacts Distribution.** The frequency histograms of values of Q visited by the LBTs were obtained according to the following procedure. Let us consider a dataset composed of trajectories obtained from MD or BF simulations. For the whole collection of trajectories we are interested in computing the fraction of native contacts, defined as

$$Q_0 = \sum_{i<j+3}^{N_R} C_{ij}^{\text{binary},0} \qquad Q(t) = \frac{1}{Q_0} \sum_{i<j+3}^{N_R} C_{ij}^{\text{binary}}(t) \tag{4}$$

where $N_R$ is the number of residues, $C^{\text{binary},0}$ is the binary contact map of the native structure, $C^{\text{binary}}(t)$ is the binary contact map at time $t$ and $Q_0$ is the total number of native contacts in the active crystal structure. The probability of finding, among all the trajectories, a particular conformation identified by a fraction of native contacts is provided by

$$p(Q) = \frac{\sum_{T=1}^{M} \left[ \sum_{\bar{Q}} \delta(Q - \bar{Q}) \right]_T}{\int dQ \sum_{T=1}^{M} \left[ \sum_{\bar{Q}} \delta(Q - \bar{Q}) \right]_T} \tag{5}$$

where the sum is extended over all M trajectories. In practice, the Dirac delta is smeared and computed by choosing a fixed bin width and counting how many conformations fall within $Q$ and $Q+dQ$ for each trajectory in the set. This discrete distribution clearly shows which regions of native contacts are highly populated and which are rarely populated, thus giving an explicit indication of possible barriers to folding and the ability of a set of trajectories to reach, for example, the native state.

**Statistical analysis of Native Contacts Distributions.** To assess the statistical significance of differences between the histograms of native contacts shown in Figure 4, we performed a two-sample Kolmogorov-Smirnov (KS) test. In this test we compared the set of Q values observed in the folding/misfolding trajectories of the different A1AT variants (e.g., WT vs. Z, Z vs. S, and so on). Because the number of LBTs differs between variants, the number of Q values used to construct the histograms also changes from variant to variant. In order to determine if the number of LBTs (and thus Q values) for each variant is statistically robust, we bootstrapped from every population a number of values of Q corresponding to 1/2 of the population in the smallest dataset which contains 29 LBTs. More specifically, we used 18192 values of Q and ran 10 bootstraps for every KS test. Our null hypothesis for the KS test assumed that the Q values obtained by simulating the folding of different A1AT variants were sampled from the same distribution. In all the KS tests, the null hypothesis was rejected with a p-value $p < 2 \times 10^{-16}$. Based on these results we conclude that all of the histograms describe statistically independent distributions. Furthermore, since the bootstrapping involved a fraction of the smallest dataset, we can conclude that the number of simulations performed for all of the variants is sufficient to provide statistically representative samples of the corresponding Q distributions. Finally, to further enforce the robustness of our results, we computed the KS test both using R (14) and SciPy (15), obtaining comparable outcomes.

As a final remark, we stress that despite the fact that the number of successful folding events is small for all A1AT variants, our main result is that the Z (Glu342Lys) mutation induces a shift of the peak of the frequency distribution to a lower (i.e. less native like) value of Q. Again, this shift is statistically significant.

**Path Similarity Analysis:** The similarity parameter measures the consistency of the temporal succession in which native contacts are formed in two given pathways. The parameter takes on values ranging from 0 for no similarity, to 1 when all native contacts form in exactly the same succession for the two trajectories.

To compute this quantity, we define a matrix *M* which describes the order of native contact formation between atoms. Given as the time of formation of the i-th native contact in the k-th trajectory, the matrix element of the k-th path is defined as:

$$M_{ij}(k) = \begin{cases} 1 & \text{if } t_{ik} < t_{jk} \\ 0 & \text{if } t_{ik} > t_{jk} \\ \frac{1}{2} & \text{if } t_{ik} = t_{jk} \end{cases} \quad (6)$$

The similarity parameter is thus defined as

$$s(k, k') = \frac{1}{N_c(N_c - 1)} \sum_{i \neq j} \delta\left(M_{ij}(k) - M_{ij}(k')\right) \quad (7)$$

where $N_C$ is the total number of native contacts. To provide a robust indication of the degree of heterogeneity of multiple trajectories, we can also compute the distribution of *s*, path similarity, over all possible pairs of trajectories, defined as

$$p(s) = \sum_{k < k'} \delta\left(s - s(k, k')\right) \quad (8)$$

The path similarity calculation allows us to quantitatively compare trajectories for a single A1AT variant, e.g., to determine similarities and differences between WT folding trajectories, and to determine whether and when the folding of A1AT mutants diverges from WT-like folding.

**Solvent Accessible Surface Area (SASA) graphs.** SASA graphs have been obtained by using the Shrake-Rupley function implemented in MDTraj 1.9.0 (16) using a probe radius of 0.144 nm and 100 sphere points.

### Supplementary References

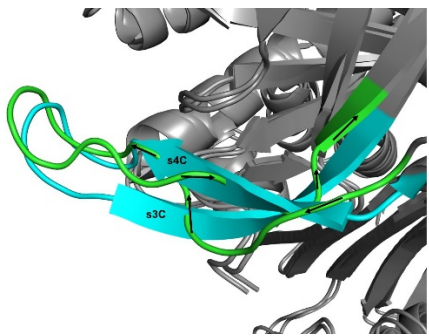

**Figure S1.** Non-native conformation in the gate region of A1AT. The crystal structure of native A1AT (PDB: 1QLP) is shown in cyan. The final structure from a representative WT least biased trajectory (LBT) that reaches the folded, native state is shown in green. As shown in the figure, in the native structure s4C passes over s3C while the reverse occurs in the successful BF WT folding trajectories. This difference in twist affects *only* the contacts at the crossover points.

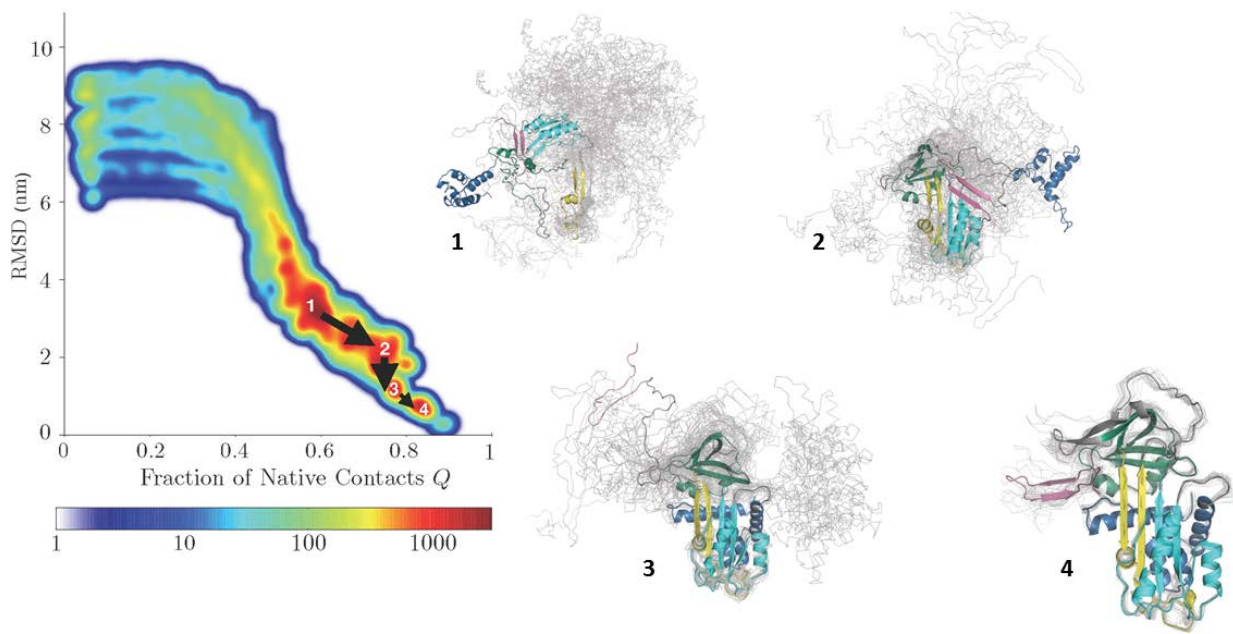

**Figure S2.** Kinetic free energy landscape and ensembles of representative structures for the Glu/Glu mutant of A1AT. See main text for details.

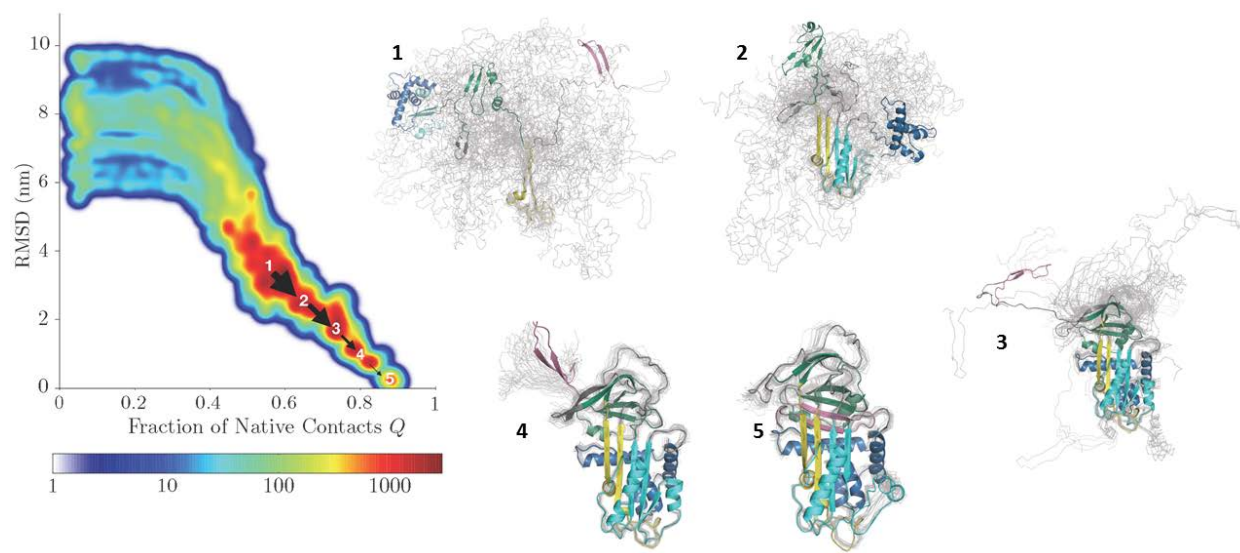

**Figure S3.** Kinetic free energy landscape and ensembles of representative structures for the Ser/Lys mutant of A1AT. See main text for details.

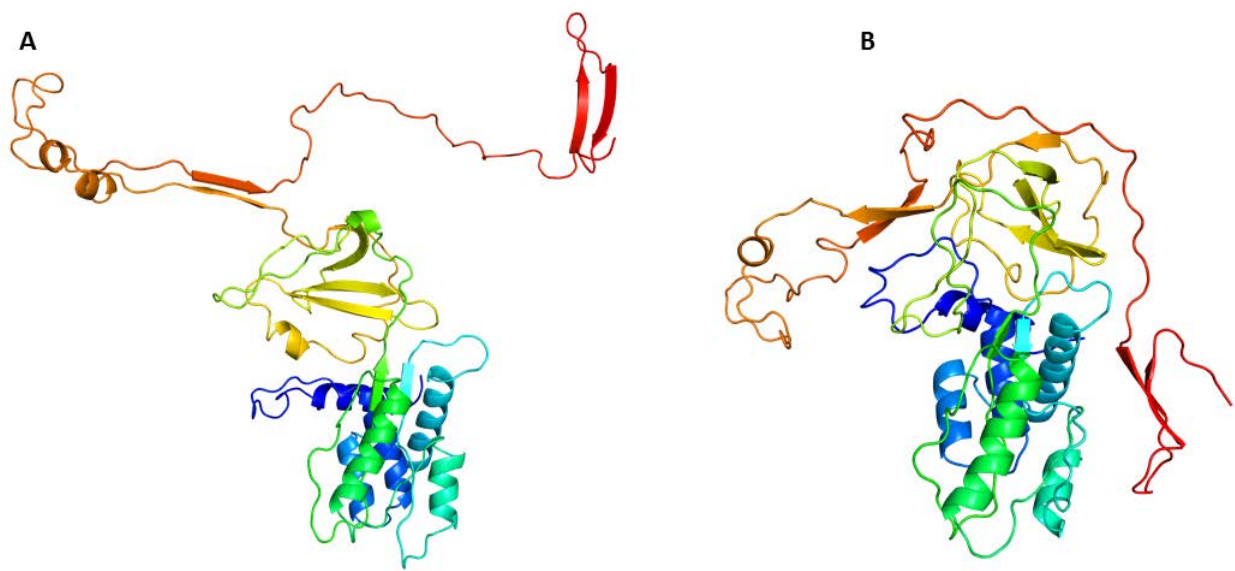

**Figure S4.** Testing the stability of a misfolded Z (Glu342Lys) A1AT structure using molecular dynamics (MD) simulations. A) The misfolded Z structure rsulting from BF simulations that was used as the initial structure for explicit solvent MD simulations. B) The structure after 200 ns of standard MD.

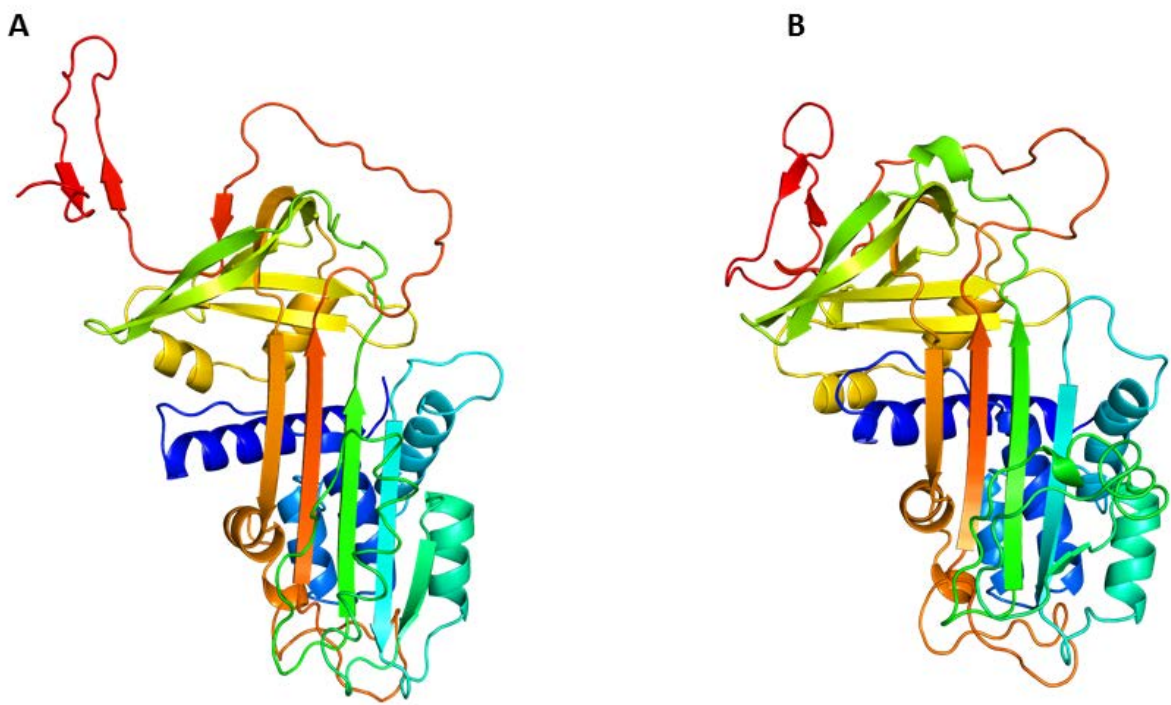

**Figure S5.** Testing the stability of a misfolded S (Glu264Val) A1AT structure using molecular dynamics (MD) simulations. A) Misfolded S structure rsulting from BF simulations that was used as the initial structure for explicit solvent MD simulations. B) The structure after 200 ns of MD.

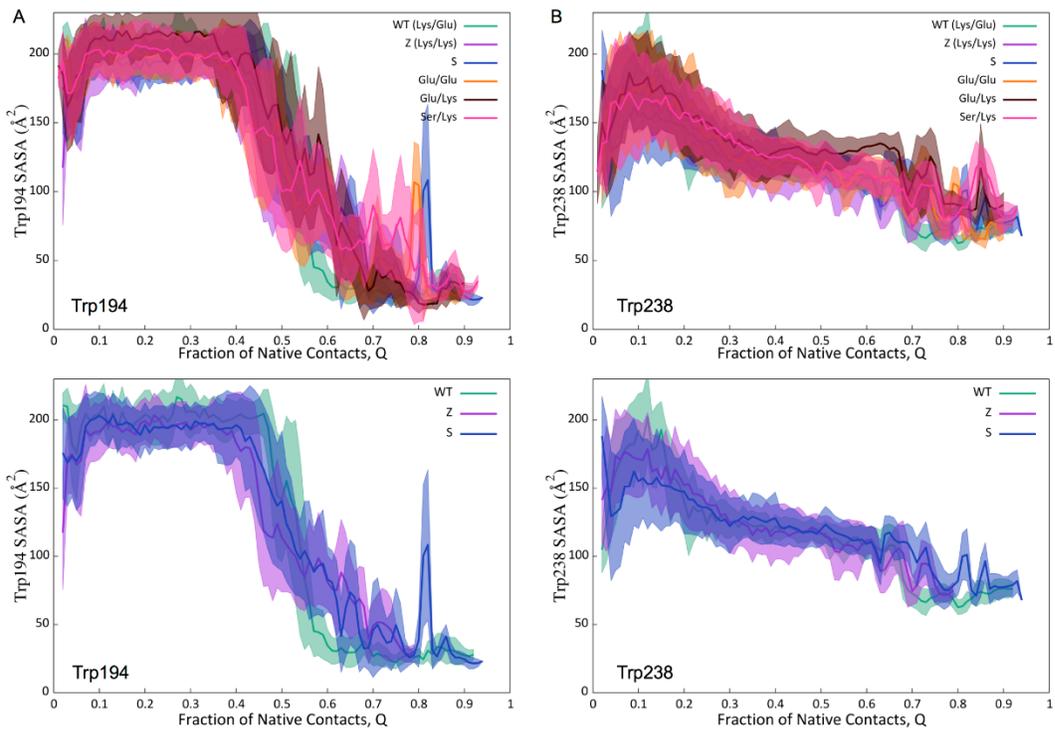

**Figure S6.** Solvent accessibility for all of the A1AT variants showing (A) Trp194 and (B) Trp238. Solid lines represent averages over all LBTs for a particular variant while shaded areas represent the standard deviations.

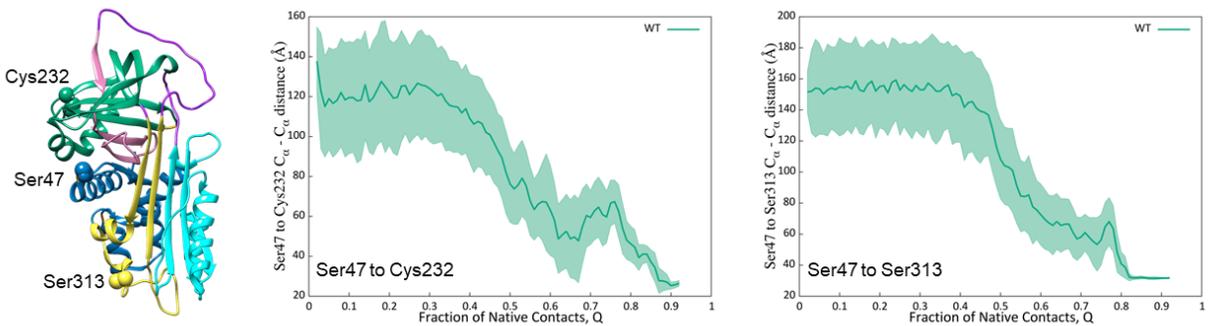

**Figure S7.** Evolution of the Ser47-Cys232 and Ser47-Ser313 distances during WT A1AT folding. Solid lines represent averages over all successful simulations while shaded areas represent the standard deviations. The locations of Ser47, Cys232 and Ser 313 in the native structure are shown on the left.